\pdfoutput=1
\documentclass[12pt]{article}
\usepackage{hyperref}
\usepackage{graphicx}
\usepackage{caption}
\usepackage[numbers,sort&compress]{natbib}
\def\msu{National Superconducting Cyclotron Laboratory and the Department of Physics and Astronomy\\ 
Michigan State University, East Lansing, MI 48824 USA}
\def\ManC{Manhattan College\\
Riverdale, NY 10471 USA
}

\def\Title#1{\begin{center} {\Large #1 } \end{center}}
\def\Author#1{\begin{center}{ \sc #1} \end{center}}
\def\Address#1{\begin{center}{ \it #1} \end{center}}

\newenvironment{Abstract}{\begin{quotation}  }{\end{quotation}}
\newenvironment{Presented}{\begin{quotation} \begin{center} 
             PRESENTED AT\end{center}\bigskip 
      \begin{center}\begin{large}}{\end{large}\end{center} \end{quotation}}
\def\Acknowledgements{\bigskip  \bigskip \begin{center} \begin{large}
             \bf ACKNOWLEDGEMENTS \end{large}\end{center}}

\def\beq{\begin{equation}}
\def\eeq#1{\label{#1}\end{equation}}
\def\eeqn{\end{equation}}

\def\beqa{\begin{eqnarray}}
\def\eeqa#1{\label{#1}\end{eqnarray}}
\def\eeqan{\end{eqnarray}}

\let\bar=\overbar

\def\Dslash{\not{\hbox{\kern-4pt $D$}}}
\def\dslash{\not{\hbox{\kern-2pt $\del$}}}

\def\msb{{\bar{\ssstyle M \kern -1pt S}}}

\begin{document}
\begin{titlepage}

\vfill
\Title{Laboratory Probes of the Neutron-Matter Equation of State}
\vfill
\Author{ M.B. Tsang, C.Y. Tsang, P. Danielewicz, and W.G. Lynch }
\Address{\msu}
\Author{F.J. Fattoyev}
\Address{\ManC}
\vfill
\begin{Abstract}
To relate constraints from nuclear physics to the tidal deformabilities of neutron stars, we construct a neutron star model that accepts input from a large collection of Skyrme density functions to calculate properties of 1.4 solar-mass neutron stars. We find that restricting this set of Skyrme to density functions that describe nuclear masses, isobaric analog states, and low energy nuclear reactions does not sufficiently restrict the predicted neutron-star radii and the tidal deformabilities. However, pressure constraints on the EoS around twice saturation density ($2\times2.74\times10^{14}g/cm^3$), obtained from high energy nucleus-nucleus collisions, does constrain predicted tidal deformabilities with uncertainties smaller than those obtained from the analysis of GW170817. We also found that the density-pressure constraint on the EoS obtained from a recent analysis of the neutron-star merger event agree very well with the density pressure constraints obtained from nuclear physics experiments published in 2002.
\end{Abstract}
\vfill
\begin{Presented}
CIPANP2018\\
Palm Springs, CA, USA, May 29 - June 3, 2018
\end{Presented}
\vfill
\end{titlepage}
\def\thefootnote{\fnsymbol{footnote}}
\setcounter{footnote}{0}

The equation of state (EoS) of nuclear matter relates temperature, pressure and density of a nuclear system. It governs both properties of nuclei and neutron stars as well as the dynamics of nucleus-nucleus collisions and that of neutron-star mergers. The EoS research in nuclear physics focuses on extrapolating the properties of neutron-rich matter from that of symmetric matter containing equal numbers of neutrons and protons \cite{r37,Hor14,r14}. This extrapolation is governed by the nuclear symmetry energy, which is defined to be the difference between the EoS of neutron matter and that of symmetric matter.  In the sub-saturation density at $0.5<\rho/\rho_0<0.7$ where $\rho_{0}$ is the saturation density, knowledge of symmetry energy is needed to predict the location of crust-core boundary in neutron stars \cite{Heb13}, crustal vibrations in Magnetars \cite{Ste09} and around $\rho/\rho_0 \approx 0.25$ to understand the supernova neutrino sphere \cite{Rob12, Hor14, Mar12}. Above saturation density, the EoS affects the fate of the neutron-star merger; whether the colliding neutron stars collapse promptly into a black hole, remain a single neutron star, or form a transient neutron star that collapses later into a black hole \cite{r4}.  \\

\begin{figure}
\centering
\includegraphics[width=0.7\linewidth]{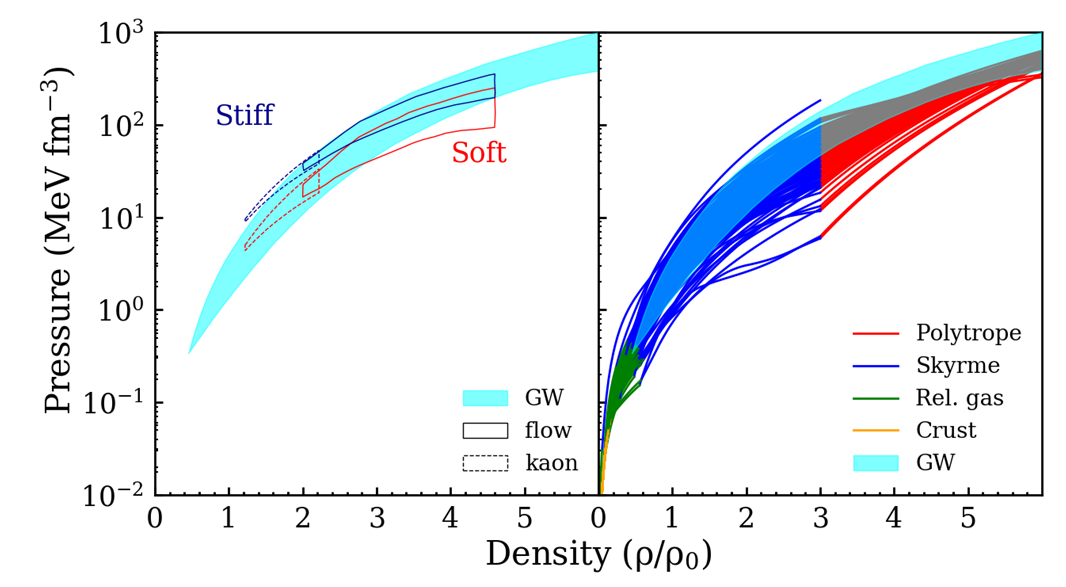} 
\caption{(Left panel) Experimental and astrophysical constraints on equation of state in pressure vs. density. The hatched region represents the GW constraint \cite{r8}, after converting the original unit for pressure of $dyn/cm^2$ to $MeV/fm^3$ and for density to units of saturation density, $\rho_0=2.74\times10^{14} g/cm^3$, to allow direct comparisons to nuclear physics constraints. Solid and dashed contours display constraints from flow measurements \cite{r37} and kaon measurements \cite{r38,r39}, respectively. The upper (labelled as stiff) and lower (labelled as soft) constraints correspond to the addition of symmetry pressure from Eqs. \ref{eq:2} and \ref{eq:3}, respectively, to the symmetric matter pressure. (Right panel) EoS used different density regions in the neutron star, the outer and inner crust region (yellow), relativistic electron Fermi gas polytropic EoS (green), Skyrme EoS (bluee) and Polytropes (red). See text for details}
\label{fig:fig1}
\end{figure}

During the inspiral phase of a neutron-star merger, the gravitational field of each neutron star induces a tidal deformation in the other \cite{r9}. The influence of the EoS of neutron stars on the gravitational wave signal during inspiral is contained in the dimensionless quantity tidal deformability, also known as tidal polarizability,$\ \Lambda=\frac{2}{3} k_2 (\frac{c^2 R}{GM})^5$, where $G$ is the gravitational constant and $k_2$ is the dimensionless Love number \cite{r5,r9}, $R$ and $M$ are the mass and radius of a neutron star. $k_2$, is sensitive to the compactness parameter $(M/R)$. As the knowledge of the mass-radius relation uniquely determines the neutron-star matter EoS \cite{r10,r11,r12,r13}, both $k_2$  and $R$ depend on the EoS. From recent analysis of the neutron star merger event GW170817, a range of EoS parameters are extracted, which are expressed in terms of the density dependence on pressure from subsaturaion to six times the saturation density as shown in the area bounded by the light blue shade areas in figure \ref{fig:fig1}. \\

In nuclear physics, the EoS of cold homogenous matter can be specified in terms of energy per nucleon of the hadronic system. Within the parabolic approximation, the EoS of cold nuclear matter can be divided into a symmetric matter contribution that is independent of the neutron-proton asymmetry and a symmetry energy term, proportional to the square of the asymmetry \cite{r14}, 

\begin{equation} \label{eq:1}
E(\rho, \delta) = E(\rho, \delta=0)+S(\rho)\delta^2
\end{equation}

where the asymmetry is defined as $\delta=(\rho_n-\rho_p)/\rho$.Here, $\rho_n, \rho_p$ and  $\rho = \rho_n+\rho_p$ are the neutron, proton and nucleon densities respectively, and $S(\rho)$ is the density dependent symmetry energy. Measurements of collective flow and kaon production in energetic nucleus-nucleus collisions have constrained the EoS for symmetric matter, $ E_(\rho, \delta=0)$ at densities up to 4.5 times saturation density \cite{r37,r38,r39,r40,r41}. In Ref \cite{r37}, the symmetric matter constraints in pressure as a function of density were determined from the measurements of transverse and elliptical flow from Au+Au collisions over a range of incident energies from 0.3 to 1.2 GeV/u. In Refs. \cite{r38,r39}, a similar constraint from $1.2\rho_0$ to $2.2\rho_0$ was obtained from the kaon measurements.  \\

To estimate the additional pressure coming from the symmetry energy, \cite{r37} uses the softest and stiffest symmetry energy functions proposed in Ref. \cite{r42} in neutron-star calculations:
\begin{equation} \label{eq:2}
S_{stiff}(\rho)=12.7 MeV\times(\frac{\rho}{\rho_0})^{\frac{2}{3}}+38MeV\times\frac{(\rho/\rho_0)^2}{1+\rho/\rho_0}
\end{equation}
\begin{equation} \label{eq:3}
S_{soft}(\rho)=12.7 MeV\times(\frac{\rho}{\rho_0})^{\frac{2}{3}}+19MeV\times(\frac{\rho}{\rho_0})^{\frac{1}{2}}
\end{equation}
For convenience, we label the functions in Eq. \ref{eq:2} and Eq. \ref{eq:3} as “stiff” and “soft”, respectively. Adding the pressure from each of these two symmetry energy functions to the pressure from the symmetric matter constraint results in two set of contours shown in the right panel of figure \ref{fig:fig1} \cite{r37}. These heavy ion constraints published decades before the neutron star merger observation agree reasonably well within the pressure-density constraints extracted in \cite{r8} represented by the light blue shaded area in figure \ref{fig:fig1}. The agreement illustrates that one can apply the knowledge gained in nuclear physics reactions to neutron star physics.  \\

To connect nuclear physics observables to neutron star observables, we will incorporate the nuclear physics density functions, $E(\rho,\delta)\ $of Eq. \ref{eq:1} into neutron star models. Most models assume that a neutron star consists of the outer crust, inner curst, outer core and inner core corresponding to different density regions where different forms of the EoS can be employed.  In our calculations, the EoS in different density regions are presented by different color curves in the left panel of figure \ref{fig:fig1}. At the lowest densities, the EoS describing both the inner and outer crust from Refs.  \cite{r18,r19,r20,r21,r22,r23,r24} are represented by yellow lines. The polytropic EoS that describes a relativistic electron fermi gas connect the curst to the inner core are represented by the green curves. The inner core region with densities between $~ 0.5\rho_0$ to $3\rho_0$ best resemble the nuclear matter environment. Here, we use the Skyrme interactions (blue curves) found in \cite{r15,r16}. Above $3\rho_0$, we use polytropic EoS of the form $K’\rho^\gamma$ to extend the EoS to the central density region of a neutron star such that it can support a maximum neutron star of 2.17 solar mass \cite{MaxMass}. \\

One advantage of using Skyrme nuclear density functions is that a large number of Skyrme interactions that describe different aspects of nuclei properties can be found and widely used in the literature \cite{r15,r16}. Skyrme interactions that generate negative or small pressure at $3\rho_0$ are eliminated. We found 189 Skyrme interactions from our collection of 207 interactions \cite{r15, r16}. The chosen EoS overlap with the gravitational constraint (light blue shaded region) quite well especially at densities below $3\rho_0$. 
 
\begin{figure}
\centering
\includegraphics[width=0.6\linewidth]{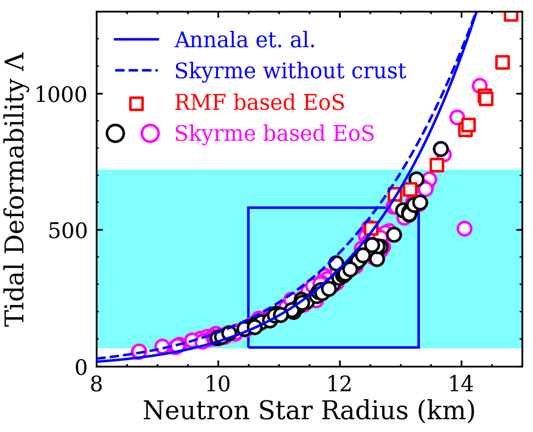}
\caption{Correlation between neutron-star tidal deformability and radii from current calculations (open circles) and from Ref. \cite{r10} (open squares). The curve is from Ref. \cite{r27} and the dashed blue curve is calculations without including crust EoS. The hatched areas represent constraints from recent GW170817 analysis \cite{r7,r8}}
\label{fig:fig2}
\end{figure}

We proceed to calculate neutron star properties such as the radii and the tidal deformability using the 189 EoS in the neutron star model. Each EoS interaction, represented by an open circle in figure \ref{fig:fig2}, makes a unique prediction for the neutron-star radius and tidal deformability. Our results are consistent with those from EoS based on relativistic mean-field interactions \cite{r10} represented by the open red squares using analogous methodology. For reference, the blue solid curve is a fitted relationship of $R$ and $\Lambda$from a “generic” neutron-star EoS \cite{r27}. Above $\Lambda> 600$, our calculations produce larger radii and deviate from the blue curve. This is due to inclusion of the crustal EoS at very low density in our calculations. If the crust EoS (the yellow and green curves) are removed and the Skyrme EoS without the beta equilibrium are extended to zero density, the blue dashed curve, a best fit curve to the results without crust is similar to the blue solid curve. Inclusion of the crust increases the neutron star radius. More detailed study of the effect of different crust EoS are underway. 

The results of the calculations are consistent with the various analysis of the GW170817 regarding the extracted tidal deformability and neutron star radius. For example, the range of the updated values of $\Lambda=70-720$ obtained in \cite{r7} is represented by the larger light blue-shaded region. The blue rectangle indicates the more restricted region of $\Lambda=70-580$ and $R = 10.5-13.3\ km$ obtained in \cite{r8}. In this work, the tighter constraint is referred simply as “GW” constraints. \\

In the past two decades, the nuclear EoS has been studied over a range of densities comparable to that found in neutron stars, in nuclear structure and reaction experiments \cite{r29,r30,r31,r32,r33,r34,r35,r36,r37,r38,r39,r40,r41}. However, in the latter context, the EoS must be extrapolated to environments where the density of neutrons greatly exceeds the density of protons. This extrapolation of the EoS to neutron–rich matter depends on $S(\rho)$ in Eq. \ref{eq:1}. Experimentally, especially at low density, effort has been made to extrapolate the symmetry energy while at high density, the experimental effort is focused more in extracting the pressure of the collision system. With our models, we can calculate both the neutron star properties such as tidal deformability and neutron star radii and nuclei properties such as neutron skins, density dependence of symmetry energy as well as density dependence of pressure as shown in figure \ref{fig:fig1}. Furthermore, the model allows us to explore other physics such as the effective masses, chemical potential equilibrium etc. in nuclear collision experiments. \\

\begin{figure}
\centering
\includegraphics[width=0.6\linewidth]{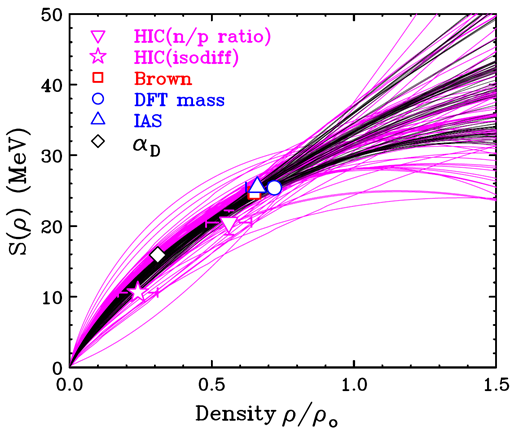}
\caption{Symmetry energy of all the Skyrmes together with experimental constraints. Black lines are Skyrmes that agrees with all low density constraints with $\chi^2 < 2$}
\label{fig:fig3}
\end{figure}

\begin{figure}
\centering
\includegraphics[width=0.6\linewidth]{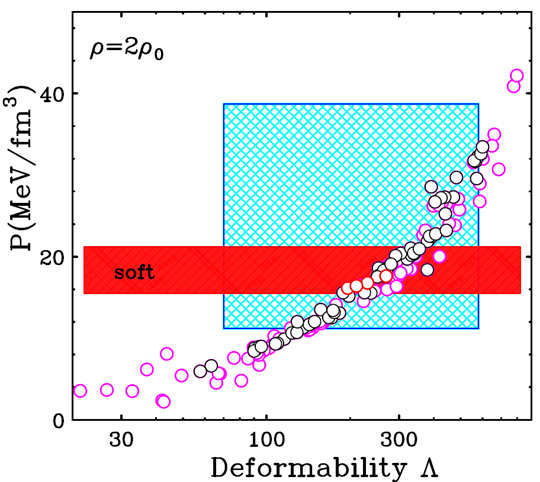}
\caption{Calculated pressure, tidal deformability and radii from the Skyrme interactions at $2\rho_0$.  The red horizontal bars represent the predicted experimental soft constraints.}
\label{fig:fig4}
\end{figure}

At sub-saturation densities, $\rho < \rho_0$, information on EoS is obtained in the context of bound nuclei and via nuclear collisions \cite{r29,r30,r31,r32,r33,r34,r35,r36}.  Figure \ref{fig:fig3} shows the symmetry energy obtained at the sensitive density of four analysis, 
1) Analyses of nuclear masses using Skyrme interactions \cite{Bro00} (red open square) and density functional theory \cite{r17} (blue open circle),
2) Analyses of isobaric analog states \cite{r15} (blue dashed contour and blue solid triangle),
3) Analyses of the electric dipole polarizability for $^{208}Pb$ \cite{Roc15,Tam11,Roc13} (green solid diamond),
4) Analyses of isospin diffusion measurements in peripheral Sn+Sn collisions at E/A=50 MeV \cite{r35,Tsa04} (magenta star). 
By evaluating the chi-squares per degree of freedom between experimental symmetry energy values obtained at various densities shown in figure \ref{fig:fig3}, 69 Skyrme functions with $\chi^2 < 2$ are plotted as black curves in figure \ref{fig:fig3} and as black open circles in Figs. \ref{fig:fig2} and \ref{fig:fig4}. These circles span over similar range of tidal polarizabilities and radii as the Skyrme EoS without any constraints suggesting that symmetry data from sub-saturation density experiments cannot adequately constraint neutron star properties. Such lack of correlation between properties of nuclei and neutron-star properties at low density should be expected. The EoS at high density reflects dominant contributions from strongly repulsive three-body interactions and few-body correlation effects that are not adequately probed by masses, isobaric analog states and other observables at low density; therefore, they remain uncertain \cite{r43}. The role of the uncertainty in three-body interactions has been elucidated in Refs. \cite{r43}. \\

The density region of $2\rho_0$ has been identified to be most sensitive to the neutron-star radii \cite{r48}. A stringent measurement on the pressure will yield a narrow range of possible deformability that can be compared to the results from neutron star merger results. As an example, the red horizontal bars in figure \ref{fig:fig4} represent the pressure ($16-20 MeV/c$) expected at $2\rho_0$ assuming the symmetry pressure from Eq. \ref{eq:3} (weak symmetry energy density dependence). In this hypothetical results, $\Lambda$ values in the range of 200-400 will be consistent with a “soft” symmetry energy. In this particular example, the uncertainties (which will shrink if the stiffness of the symmetry pressure increases) are smaller than the uncertainty of the GW constraint. The expected improvement in the HIC constraints will allow us to understand what these combined constraints imply about the nature of strongly interacting matter. \\

In summary, while symmetry energy data extracted at low density do not correlate strongly with neutron-star properties, heavy ion collision experiments testing twice the normal nuclear matter density may provide tighter constraints on the tidal deformability and the corresponding neutron-star radii.

\Acknowledgements
This work was partly supported by the US National Science Foundation under Grant PHY-1565546 and by the U.S. Department of Energy (Office of Science) under Grants DE-SC0014530, DE-NA0002923.

\bibliography{xampl}

\begin{thebibliography}{99}



\bibitem{r37}
P. Danielewicz, R. Lacey, and W. G. Lynch, Science 298, 1592 (2002).    
\bibitem{Hor14}
C.J. Horowitz, E.F. Brown, Y. Kim, W.G. Lynch, R. Michaels, A. Ono, J. Piekarewicz, M.B.  Tsang, and H.H. Wolter, J. Phys. G 41, 093001 (2014).
\bibitem{r14}
B-A Li, Lie-Wen Chen, Che Ming Ko, Phys. Rep. 464, 113-281 (2008).
\bibitem{Heb13}
K. Hebeler, J. M. Lattimer, C. J. Pethick and A. Schwenk, Astrophys. J. 773, 11 (2013).
\bibitem{Ste09}
A. W. Steiner and A. L. Watts, Phys. Rev. Lett. 103, 181101 (2009).
\bibitem{Rob12}
L. F. Roberts, Sanjay Reddy, and Gang Shen, Phys. Rev. C 86, 065803 (2012).
\bibitem{Mar12}
G. Martinez-Pinedo, T. Fischer, A. Lohs, and L. Huther, Phys. Rev. Lett. 109, 251104 (2012).
\bibitem{r4}
A. Bauswein, O. Just, H-T Janka, N. Stergioulas, The Astrophysical Journal Letters, 850:L34 (2017).
\bibitem{r9}
K. Chatziioannou, C-J Haster, A. Zimmerman, Phys. Rev. D 97,104036 (2018). 
\bibitem{r5}
B.P. Abbott et al., Phys. Rev. Lett. 119, 161101 (2017). 
\bibitem{r10}
F. J. Fattoyev, J. Piekarewicz, C. J. Horowitz, Phys. Rev. Lett. 120, 172702 (2018).
\bibitem{r11}
S. Postnikov, M. Prakash and James Lattimer, Phys.Rev.D 82, 024016, (2010).
\bibitem{r12}
J. M. Lattimer and M. Prakash, Phys. Rep. 621, 127-164 (2016).
\bibitem{r13}
L. Lindblom, Astrophys. J. 398, 569 (1992).  
\bibitem{r38}
C. Fuchs, Nucl. Phys. 56, 1-103 (2006). 
\bibitem{r39}
W.G. Lynch, M.B. Tsang, Y. Zhang, P. Danielewicz, M. Famiano, Z. Li, A.W. Steiner; Prog. Part. Nucl. Phys. 62, 427-432 (2009).  
\bibitem{r40}
A. LeFèvre, Y. Leifels, W. Reisdorf, J. Aichelin, Ch. Hartnack, Nucl. Phys. A 945, 112 (20).
\bibitem{r41}
P. Russotto et al. Phys. Rev. C 94, 034608 (2016).
\bibitem{r42}
M. Prakash, T. L. Ainsworth and J.M. Lattimer, Phys. Rev. Lett. 61, 2518 (1988).
\bibitem{r8}
B. P. Abbott, et al., Phy. Rev. Lett. 121, 161101 (2018). 
\bibitem{r18}
G. Baym, C. Pethick, P. Sutherland, Astrophys J., 170:299-137 (1971).
\bibitem{r19}
C. P. Lorenz, D. G. Ravenhall, and C. J. Pethick, Phys. Rev. Lett. 70, 379 (1993).
\bibitem{r20}
A. S. Schneider, C. J. Horowitz, J. Hughto, D. K. Berry, Phys. Rev. C88, 065807 (2013).
\bibitem{r21}
A. S. Schneider, D. K. Berry, C. M. Briggs, M. E. Caplan, C. J. Horowitz, Phys. Rev. C90, 055805 (2014). 
\bibitem{r22}
D. K. Berry, M. E. Caplan, C. J. Horowitz, G. Huber, A. S. Schneider, Phys. Rev. C94, 055801 (2016).
\bibitem{r23}
F. J. Fattoyev, C. J. Horowitz, B. Schuetrumpf, Phys. Rev. C 95, 055804 (2017). 
\bibitem{r24}
J.W. Negele, D. Vautherin, Nuclear Physics A 207:298-320 (1973). 
\bibitem{r15}
P. Danielewicz, Pardeep Singh, Jenny Lee, Nucl. Phys. A 958, 147-186 (2017).
\bibitem{r16}
M. Dutra, O. Lourenco, J.S. Sa Martins, A. Delfino, J.R. Stone, P.D. Stevenson, Phys. Rev. C 85, 035201 (2012). 
\bibitem{MaxMass}
B. Metzger, As B. Margalit and B. D. Metzger, Astrophys. J. Lett. 850, L19 (2017).
\bibitem{r27}
E. Annala, T. Gorda, A. Kurkela, A. Vuorinen, Phys. Rev. Lett. 120, 172703 (2018). 
\bibitem{r7}
B. P. Abbott, et al., arXiv:1805.11579 (2018). 
\bibitem{r29}
W.G. Lynch, M.B. Tsang, arXiv:1805.10757  
\bibitem{r30}
B.A. Brown, Phys. Rev. Lett. 111, 232502 (2013).
\bibitem{r31}
M. Kortelainen, T. Lesinski, J. Moré, W. Nazarewicz, J. Sarich, N. Schunck, M. V. Stoitsov, and S. Wild, Phys. Rev. C 82, 024313 (2010).  
\bibitem{r32}
D.H. Youngblood, H.L. Clark, Y.W. Lui, Phys. Rev. Lett. 82, 691 (1999).
\bibitem{r33}
A. Tamii, et al., Phys. Rev. Lett. 107, 062502 (2011). 
\bibitem{r34}
Zhen Zhang and Lie-Wen Chen, Phys. Rev. C 92, 031301(R) (2015).    
\bibitem{r35}
M.B. Tsang, Y.X. Zhang, P. Danielewicz, M. Famiano, Z.X. Li, W.G. Lynch and A.W. Steiner, Phys. Rev. Lett. 102, 122701 (2009).
\bibitem{r36}
D.D.S. Coupland, M. Youngs, Z. Chajecki, W.G. Lynch, M. B. Tsang, Y.X. Zhang, M.A. Famiano, T. Ghosh, et al. Phys. Rev. C 94 011601(R) (2016).
\bibitem{Bro00}
B.A. Brown, Phys. Rev. Lett. 85, 5296 (2000).
\bibitem{r17}
F. J. Fattoyev, J. Carvajal, W. G. Newton, B-A Li, Phys. Rev. C 87, 015806 (2013). 
\bibitem{Roc15}
X. Roca-Maza, X. Viñas, M. Centelles, B. K. Agrawal, G. Colò, N. Paar, J. Piekarewicz, and D. Vretenar, Phys. Rev. C 92, 064304 (2015).
\bibitem{Tam11}
A. Tamii, I. Poltoratska, P. vonNeumann-Cosel, Y. Fujita, T. Adachi, C. A. Bertulani, J. Carter, M. Dozono, H. Fujita, K. Fujita, K. Hatanaka, D. Ishikawa, M. Itoh, T. Kawabata, Y. Kalmykov, A. M. Krumbholz, E. Litvinova, H. Matsubara, K. Nakanishi, R. Neveling, H. Okamura, H. J. Ong, B. Ozel-Tashenov, V. Y. Ponomarev, A. Richter, B. Rubio, H. Sakaguchi, Y. Sakemi, Y. Sasamoto, Y. Shimbara, Y. Shimizu, F. D. Smit, T. Suzuki, Y. Tameshige, J. Wambach, R. Yamada, M. Yosoi, and J. Zenihiro, Phys. Rev. Lett. 107, 062502 (2011).
\bibitem{Roc13}
X. Roca-Maza, M. Brenna, and G. Colò, M. Centelles and X. Viñas, B. K. Agrawal, N. Paar and D. Vretenar, J. Piekarewicz, Phys. Rev. C 88, 024316 (2013).
\bibitem{Tsa04}
M. B. Tsang et al., Phys. Rev. Lett. 92, 062701 (2004).
\bibitem{r43}
A.W. Steiner, J.M. Lattimer, E.F. Brown, Astrophys. J. Lett. 765, L5 (2013).
\bibitem{r48}
S. Gandolfi, J. Carlson, S. Reddy, Phys. Rev. C 85, 032801(R) (2012).


\end{thebibliography}
 
\end{document}